\documentclass[twoside,aps,superscriptaddress,pre,showpacs,floatfix]{revtex4}
\usepackage[dvips]{color}

\usepackage{amsmath,amssymb,graphicx,latexsym}
\usepackage[T1]{fontenc}

\newcommand{\bq}{\begin{equation}} 
\newcommand{\eq}{\end{equation}}   
\newcommand{\refm}[1]{(\ref{#1})}  

\newcommand{\mitt}[1]{\left\langle #1 \right\rangle}

\newcommand{\om}{\omega}

\newcommand{\vH}{\mathbf{H}}
\newcommand{\N}{\mathbf{N}}

\newcommand{\m}{\mathbf{m}}

\newcommand{\e}{\mathbf{e}}

\newcommand{\di}{\partial}

\newcommand{\vS}{\boldsymbol{S}}
\newcommand{\Be}[1]{B\left( #1 \right)}

\newcommand{\vu}{\boldsymbol{u}}
\newcommand{\kb}{k_ {\text{B}}}

\begin{document}
\title{Role of interactions in ferrofluid thermal ratchets}
\author{Volker Becker}
\email{becker@theorie.physik.uni-oldenburg.de}
\author{Andreas Engel}
\email{engel@theorie.physik.uni-oldenburg.de} 
\affiliation{Institut f\"ur Physik, Carl-von-Ossietzky-Universtit\"at,
     26111 Oldenburg, Germany }
\begin{abstract}
Orientational fluctuations of colloidal particles with magnetic
moments may be rectified with the help of external magnetic fields
with suitably chosen time dependence. As a result a noise-driven
rotation of particles occurs giving rise to a macroscopic torque per
volume of the carrier liquid. We investigate the influence of mutual
interactions between the particles on this ratchet effect by  
studying a model system with mean-field interactions. The stochastic
dynamics may be described by a nonlinear Fokker-Planck equation for
the collective orientation of the particles which we solve
approximately by using the effective field method. We determine an
interval for the ratio between coupling strength and noise intensity
for which a self-sustained rectification of fluctuations becomes
possible. The ratchet effect then operates under conditions for
which it were impossible in the absence of interactions.  
\end{abstract}
\maketitle


\section{Introduction}
Thermal ratchets or Brownian motors are devices that are able to
rectify thermal fluctuations and thereby extract
directed motion from irregular microscopic chaos. Banished from
equilibrium by the second law of thermodynamics they may nevertheless
be found in various non-equilibrium situations (for a review see
\cite{Reimann02}). Having being used for a long time as thought models
to highlight some subtle points in the foundations of statistical
mechanics \cite{HaRe} they became recently objects of more practical
interest due to their possible relevance for biological 
transport mechanisms \cite{Mag,JulicherAP97} and potential
applications in nano-technology \cite{RSAP,OuBo,Lin}. 

Among the many devices that have been proposed as Brownian motors 
\cite{OuBo} those based on ferrofluids \cite{EngelR04,EngelMRJ03} are
of particular interest since they allow a rather direct observation
of the ratchet effect on the macroscopic level. Ferrofluids are
suspensions of ferromagnetic nano-particles in carrier liquids like
water or oil combining the hydrodynamic behaviour of Newtonian fluids
with the magnetic properties of superparamagnets \cite{Ros}. The
orientational Brownian motion of the ferromagnetic grains can be
rectified with the help of a suitably chosen time-dependent magnetic
field. Due to the viscous coupling between the rotation of the
magnetic grains and the local vorticity of the hydrodynamic flow the
angular momentum produced by the ratchet is transferred to the carrier
liquid and may be detected as hydrodynamic torque per fluid volume
\cite{EngelR04}.  

Although the interactions between the ferromagnetic particles
mediated by the carrier liquid are of crucial importance for the
macroscopic manifestation of the ratchet effect it was neglected in
the theoretical modeling done so far \cite{EngelR04,EngelMRJ03}. In
the present paper we want to remedy of this drawback and to elucidate
the influence of direct (dipole-dipole) and indirect (hydrodynamical)
interactions between the ferrofluid particles on the ratchet effect in
ferrofluids. This will be done using a simplified mean-field model of
interactions which allows to analytically discuss some of the
qualitative changes that occur.

General aspects of the interplay between the ratchet effect and
interactions were already discussed in \cite{VandenBroeckBRL00} and
\cite{Reimann02}. In particular it was shown that under certain
conditions the interactions may give rise to a spontaneous breakdown
of symmetries that formerly inhibited the ratchet effect. The
relevance of interactions in biological applications was investigated
in \cite{JulicherAP97}.

Here we will find by studying a concrete, experimentally accessible
example of a ratchet that depending on the ratio between coupling
strength and noise intensity qualitatively different regimes are
possible. In some of these regimes a self-sustained ratchet effect may occur
under circumstances for which no such effect would show up in the
absence of interactions. The discussion will be centered around a non-linear
Fokker-Planck equation for the stochastic dynamics of the collective
orientation of the ferrofluid particles \cite{DesaiZ78} which is
approximately solved by adapting the well-known effective-field method
from the theory of ferrofluids \cite{RaikherS94} to the analytical
investigation of the ratchet effect in these systems.

The paper is organized as follows. In section II we introduce the
model and the basic equations and derive the non-linear Fokker-Planck
equation for the collective orientation of the ferromagnetic
particles. Section III contains some numerical solutions of this
Fokker-Planck equation providing first insight into the
different regimes that are possible in the system. In section IV we
introduce the effective-field method and show how it may be adapted to
the present situation. Using this method we will characterize in
Section V the different regimes found in section III, determine 
their stability and discuss the different manifestations of the
ratchet effect. Finally, section VI contains our conclusions. 


\section{The Model}
We consider the overdamped dynamics of $N$ identical spherical
particles with magnetic moments  $\m_i, i=1,...,N$, dispersed in a
Newtonian liquid with viscosity $\eta$. The particles are under the
influence of a time-dependent external magnetic field with constant
$x$-component and time-periodic $y$-component
\bq
    \vH= \left(H_x,H_y(t),0 \right)\;,\;\;
    H_y\left(t+\frac{2\pi}{\Omega}\right)=H_y(t)\; .  \label{timedep} 
\eq
A suitable example for the time dependence of the $y$-component is
\cite{EngelMRJ03} 
\bq\label{foft}
        H_y(t)=\alpha \cos(\Omega t) + \beta \cos(2\Omega t+\delta)\; ,
\eq
with the scalar parameters $\alpha, \beta$, and $\delta$. 

The change of orientation $ \vu_i = \m_i/m_i$  of particle $i$ is given
by  
\bq
        \di_t \vu_i = \boldsymbol{\om_i} \times \vu_i \; ,
\eq
where $\boldsymbol{\om_i}$ denotes the angular velocity the particle. 

The particle orientations tend to align with the magnetic
field. Since the latter is confined to the $x$-$y$ plane the average
orientations will have zero $z$-components and the main effects will
show up in the time-dependence of their $x$- and $y$-components. To
keep the analysis simple we will therefore only consider the dynamics
in the $x$-$y$ plane and put $u_{i,z} \equiv 0$ for all $i$. It is
then convenient to parametrize the orientations in the form
\bq
        \vu_i=\left( \begin{matrix} \cos \phi_i \cr \sin \phi_i \cr 0
          \end{matrix} \right) \; .
\eq
In the overdamped limit the angular velocities $\boldsymbol{\om_i}$
are determined by the balance of torques  
\bq\label{baltorque}
        0 = \N_{magn,i}+\N_{visc,i}+\N_{stoch,i}+ \N_{int,i} \; .
\eq
The different contributions in this equation denote the magnetic
torque due to the interaction with the external magnetic field
\bq
    \N_{magn,i}= m \, \vu_i \times \vH=m \big(H_x \sin\phi_i -
          H_y(t) \cos\phi_i\big) \e_z \; , 
\eq
the viscous and stochastic torques
\bq
        \N_{visc,i}=-6 \eta V \boldsymbol{\omega_i}=-6 \eta V \di_t
        \phi\; \e_z \; ,
\eq
and
\bq
    \N_{stoch,i}=\sqrt{12 \eta V \kb T} \, \xi_i(t)\, \e_z\; ,
\eq
respectively describing the interaction with the carrier fluid, and
the torque arising from the interaction between the particles 
\bq\label{defNint}
        \N_{int,i}=\frac{K}{N} \sum_{j=1}^N \vu_j \times
        \vu_i=\frac{K}{N} \sum_{j=1}^N \sin(\phi_i - \phi_j)\, \e_z \;
        . 
\eq
In the above expressions $\kb$ denotes Boltzmann's constant, $T$
temperature, and $V$ the volume of the particles. The $ \xi_i(t)$ are
identical and independent Gaussian noise sources with zero mean and
correlation  
\bq
        \mitt{\xi_{i}(t) \xi_{j}(t')} = 
         \delta_{ij}\, \delta (t-t') \; .
\eq
The interaction term \refm{defNint} is of mean-field type as e.g. also
advocated in \cite{VandenbroeckPT94,Reimann02}. Although the direct
magnetic dipole-dipole  
interaction as well as the indirect hydrodynamic interaction between the
particles are much more complicated than the simple
assumption \refm{defNint} several interesting implications of
interactions will become apparent already in this mean-field
description.  

>From \refm{baltorque} we find  the following set of Langevin equations
describing the stochastic dynamics of the system
\bq
\di_t \phi_i = \frac{1}{6 \eta V} \Big(m H_x+\frac{K}{N} \sum_{j=1}^{N}
  \cos \phi_j \Big) \sin \phi_i - \frac{1}{6 \eta V } \Big(m 
  H_y(t)+\frac{K}{N} \sum_{j=i}^N \sin \phi_j \Big) \cos \phi_i +
\sqrt{2D} \, \xi_i(t)  
\eq
with the diffusion coefficient $D$ given by 
\bq 
   D:= \frac{\kb T}{6 \eta V} \; .
\eq
It is convenient to introduce dimensionless quantities by performing
the rescalings 
$t \rightarrow t /  \Omega $, $\vH \rightarrow 6 \eta V \Omega / m \,\vH
$, $D \rightarrow  \Omega^2 D $ and $K \rightarrow 6 \eta V \Omega \,
K$ to obtain 
\bq
\di_t \phi_i = \Big(H_x+\frac{K}{N} \sum_{j=1}^{N} \cos \phi_j
\Big) \sin \phi_i - \Big( H_y(t)+\frac{K}{N} \sum_{j=i}^N \sin
  \phi_j \Big) \cos \phi_i + \sqrt{2D}\, \xi_i(t)\; .
\label{Langevingleichung} 
\eq
Equivalent to this set of Langevin equations is a Fokker-Planck
equation for the joint probability distribution $W(\phi_i,t)$ of the
particle orientations. 

As usual in a mean-field model we are interested in the 
limit $N\to \infty$. It is then useful to introduce the distribution
function for particle orientations 
\bq\label{defP}
   P(\phi,t)=\frac{1}{N} \sum_{i=1}^N \delta \left( \phi - \phi_i(t)
   \right) \; .   
\eq
As discussed in detail in
\cite{DesaiZ78,Dawson83,Bonilla87,StrogatzM91} $P(\phi,t)$ becomes for
$N \to\infty $ {\em independent} of the specific realization of the
noise $\xi_i(t)$ (self-averaging property). Consequently the same
holds true for the  
{\em collective orientation} of the particles defined by  
\bq\label{sundu}
   \boldsymbol{S}(t)=\frac{1}{N} \sum_{i=1}^N \vu_i(t) 
        = \int_0^{2\pi} d\phi \;\vu\, P(\phi,t)=
  \left( \begin{matrix} \mitt{\cos \phi} \cr \mitt{\sin \phi} \cr 0
          \end{matrix} \right)\; ,
\eq
where  $\mitt.$  denotes the average with $P(\phi,t)$. 
We therefore get from \refm{Langevingleichung} $N$ {\it decoupled} 
Langevin equations of the form
\bq \label{mflangevin}
  \di_t \phi_i= \big( H_x+K S_x(t) \big) \sin \phi_i - \big( H_y(t)+K S_y(t)
  \big) \cos \phi_i+\sqrt{2D} \, \xi_i(t) \; . 
\eq   
The solution $W(\phi_i,t)$ of the equivalent Fokker-Planck equation 
now factorizes, $W(\phi_i,t)=\prod_{i=1}^N w(\phi_i,t)$,
with $w(\phi_i,t)$ obeying the {\em single-particle} Fokker-Planck equation 
\cite{Coffey98,MartsenyukRS73}
\bq
\di_t w(\phi,t) = \di_\phi \Big( w(\phi,t) \di_\phi U(\phi,t)
\Big) + D \, \di_\phi^2 w(\phi,t) \label{FPE} \; .
\eq
The potential $U$ is defined by 
\bq
    U(\phi,t)=-\vu \cdot (\vH(t) + K \vS(t))=-(H_x+K S_x(t)) \cos \phi - 
     (H_y(t)+K S_y(t)) \sin \phi  \label{potential}\; , 
\eq
and depends parametrically on $\vH(t)$ and $\vS(t)$. 

We may now average \refm{defP} over the realizations of the $\phi_i(t)$
to find that $P$ obeys the same equation as $w$, 
\bq\label{NFPE}
\di_t P(\phi,t) = \di_\phi \big( P(\phi,t) \, 
   \di_\phi U(\phi,t;\vH(t),\vS(t)) \big) + D\, \di_\phi^2 P(\phi,t) \; .
\eq
This equation is closed by \refm{sundu}. As characteristic for a
mean-field system we have thus reduced the dynamics of $N\to\infty$
coupled degrees of freedom to the dynamics of a single degree of
freedom in an augmented field 
\begin{equation}
  \label{mfhfeld}
  \vH_{mf}(t)=\vH(t)+K \vS(t)
\end{equation}
to be determined {\em self-consistently}. For a related application of
mean-field techniques to ferrofluids see also \cite{LuEm}. 

\begin{figure}
\begin{center}
\includegraphics[width=0.5\textwidth]{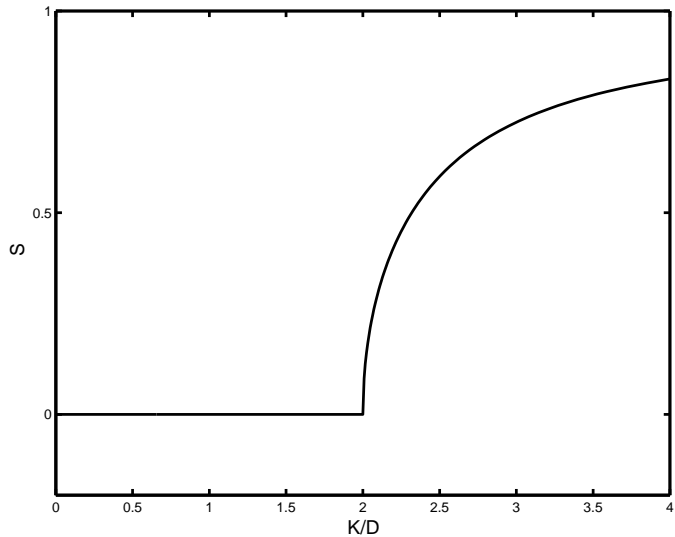}
\end{center}
\caption{Modulus $S$ of the collective orientation $\vS$ defined in
  \refm{sundu} as function of the ratio $K/D$ between interaction and disorder
strength. In the disordered phase, $K/D<2$, $S$ is identically zero,
for $K/D>2$ a spontaneous ordering takes place in the mean-field
model.} \label{ggm} 
\end{figure}

For a static field, $\vH(t)\equiv \vH$, eq.~\refm{NFPE} admits the
stationary solution  
\bq \label{Peq}
    P_{eq}(\phi) = \frac{1}{\displaystyle 2 \pi I_0(\frac{|\vH_{mf}|}{D})}
  \exp \left( \frac{H_{mf,x}\cos \phi + H_{mf,y} \sin \phi}{D} \right)
  \; . 
\eq
Here and in the following $I_n(x)$ denotes the $n$-th Bessel function of
complex arguments. In the absence of an external field, $\vH=0$, the
self-consistency condition \refm{sundu} takes the form 
\bq\label{Sav}
    \vS= \left( \begin{matrix} \mitt{\cos \phi} \cr \mitt{\sin
     \phi} \end{matrix} \right)= \Be{\frac{K S}{D}} \frac{\vS}{S} \; ,
\eq
where we introduced the function 
\bq\label{defBx}
        B(x)=\frac{I_1(x)}{I_0(x)} \; ,
\eq
which is a monotonously increasing and satisfies 
\begin{align}\label{propB1}
  B(x)&\sim \frac{x}{2} \qquad\text{for}\qquad x\to 0\\
  B(x)&\to 1 \qquad\text{for}\qquad x\to \infty\; .\label{propB2}
\end{align}
Eq.~\refm{Sav} coincides with the self-consistent equation for the
ferromagnetic mean-field $x$-$y$ model. The dependence of the modulus
$S$ of the collective orientation $\vS$ on the ratio between
interaction and fluctuation strengths is shown in Fig.~\ref{ggm}. Note
that the direction of $\vS$ in the ordered phase is arbitrary. Note
also that for short range interactions the situation is rather different
\cite{MerminWagner,KosterlitzThouless}. 

The central quantity of interest in connection with the ratchet effect
in the present system is the time and ensemble averaged torque
transferred from the magnetic field to the particles \cite{EngelMRJ03,EngelR04}
\bq
 \overline{\mitt{\N}} = \lim_{T\to\infty} \frac{1}{T}
        \int_0^T dt \mitt{\vu} \times  \vH 
  = -\lim_{T\to\infty} \frac{1}{T}
        \int_0^T dt\mitt{\di_t \phi} \e_z\; ,
\eq
where the second equality follows from \refm{mflangevin}. 

In the absence of interactions between the particles the necessary
conditions for the ratchet effect to operate, i.e. for 
$\overline{\mitt{N_z}}\neq 0$, have been discussed in detail in 
\cite{EngelR04,beckerE05}. In particular it was shown by a symmetry
argument that $H_x=0$ implies $\overline{\mitt{N_z}}=0$. In the present
mean-field model the role of $\vH$ is played by $\vH_{mf}=\vH+\vS$ and
it might therefore be possible to find $\overline{\mitt{N_z}}\neq 0$
whenever $H_{mf,x}\neq 0$. However, the latter condition may be
fulfilled by $H_x=0$ and $S_x\neq 0$. Hence we may suspect
that rectification of fluctuations may take place in the interacting
system even under conditions for which it would be impossible in a
system without interactions. Note that there is no trivial mapping
between the two cases since $\vH_{mf,x}$ is time dependent and has to
be determined self-consistently. Nevertheless the detailed
investigations discussed below will show that the interactions between
the particles may indeed give rise to a non-zero value of $S_x$ which
in turn may drive the ratchet effect in the system even if $H_x=0$.
  

\section{Numerical solution of the Fokker-Planck Equation} \label{NUM}
In the present section we investigate some general features of the
dynamics of the system under consideration with the help of a
numerical solution of our central equation \refm{NFPE}. To this end we
expand $P(\phi,t)$ in Fourier modes with respect to $\phi$
\bq\label{Fourier}
        P(\phi,t) =\sum_{n=-\infty}^{\infty} a_n(t) \exp(i n \phi)\; 
\eq
with time-dependent complex expansion coefficients $a_n(t)$. 
From \refm{sundu} we then have 
\begin{align}
        S_x(t) &= 2 \pi \, \Re a_1(t) \\
        S_y(t) &= -2 \pi \, \Im a_1(t) \; , 
\end{align}
whereas the average torque is given by
\bq\label{Nnum}
   \overline{\mitt{N_z}}= \overline{H_x \Im a_1 + H_y \Re a_1}
\eq
Using the Fourier expansion of $P(\phi,t)$ in \refm{NFPE} we
obtain an infinite system of coupled ordinary differential equations
for the coefficients $a_n(t)$ of the form  
\bq \label{entwicklung}
   (\di_t + D n^2) a_n(t) =\frac{n}{2} (g(t) a_{n-1} - g^*(t)
   a_{n+1} )\; .
\eq
where 
\bq
       g(t) = H_x - i H_y + 2 \pi K a_1(t)  \; .
\eq
Starting with 
\bq
        a_0=\frac{1}{2 \pi}
\eq
as required by normalization of $P(\phi,t)$ we may solve the
hierarchy of equations \refm{entwicklung} iteratively up to 
some value $n_{max}$ of $n$. Using different values for $n_{max}$ we
have found that $n_{max}=10$ is sufficient to get accurate numerical
results. 

\begin{figure}
\begin{center}
\includegraphics[width=0.6\textwidth]{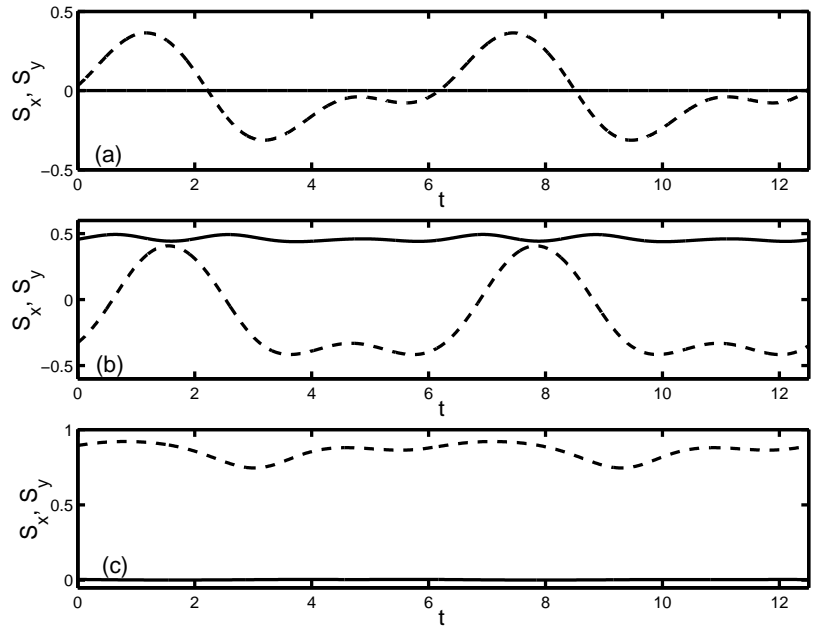}
\end{center}
\caption{Typical time dependence of the collective orientation $\vS$
  in the three different regimes described in the text. Shown are $S_x(t)$
  (dashed lines) and $S_y(t)$ (solid lines). The parameter values
  are $H_x=0$, $\alpha=\beta=1/\sqrt{2}$, and $K=3$.
  (a) Regime 1: $D=2.4$, $\overline{\mitt{N_z}}=0$ (b) Regime 2:
  $D=1.2$, $\overline{\mitt{N_z}}=3.5\,10^{-5}$ (c) Regime 3: $ D=0.6$
  $\overline{\mitt{N_z}}=0$.} 
\label{regimes}
\end{figure}

As discussed at the end of the previous section the case $H_x=0$ is of
particular interest. Depending on the ratio between coupling strength
and noise intensity we find in this case three different regimes
which are characterized by the time dependence of the average
orientation $\vS(t)$ as shown in Fig.~\ref{regimes}. In
the first regime, $K/D<2$, we find $S_x(t)\equiv 0$ and
$S_y(t)$ changes sign during one period of $H_y(t)$. In the
second regime, $2<K/D\lesssim3.7$, $S_x$ acquires non-zero values 
whereas the behaviour of $S_y(t)$ is qualitatively similar to
the first case. In the third regime, $3.7\lesssim K/D$, we have again
$S_x\equiv 0$ but now $S_y(t)$ is positive for all $t$. Similar
regimes have been discussed also for the spherical model
\cite{DharT92} and the anisotropic $x$-$y$ model \cite{YasuiTYF02} in
time dependent external fields. 

\begin{figure}
\begin{center}
\includegraphics[width=0.6\textwidth]{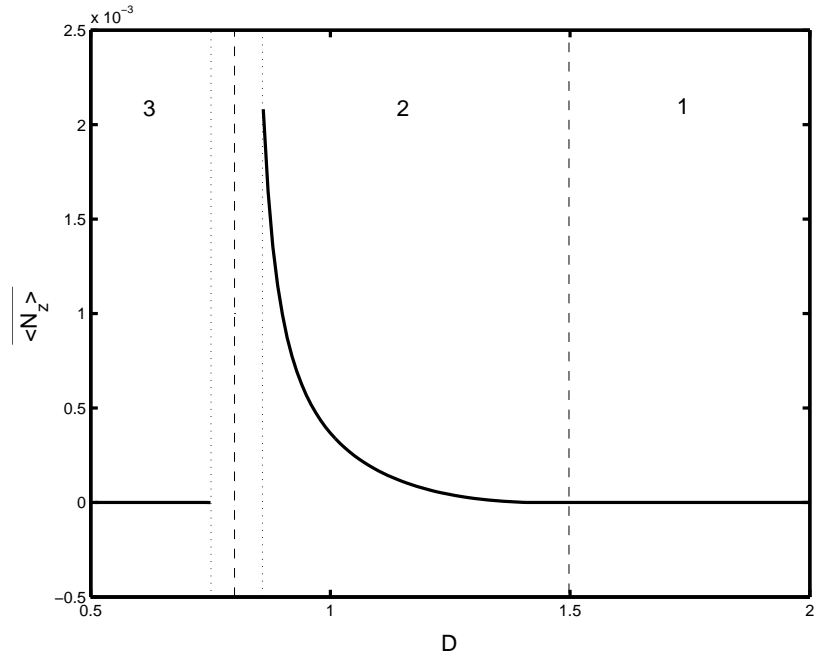}
\end{center}
\caption{Numerical results for the time averaged torque
  $\overline{\mitt{N_z}}$ for $H_x=0$ as a function of the disorder 
  strength $D$ for $K=3, \alpha=\beta=1/\sqrt{2}, H_x=0$. The vertical
  dashed lines divide the different regimes from each other ($1.5<D$, 
  regime 1; $0.8<D<1.5$, regime 2; $D<0.8$, regime 3). In the
  interval indicated by the dotted lines,  $0.75<D<0.86$, no
  asymptotic value of $\overline{\mitt{N_z}}$ could be obtained after
  $200$ periods of driving. \label{tourqehx0}} 
\end{figure}

As discussed in the previous section the symmetry analysis performed
in \cite{EngelR04} implies that for $H_x=0$ and $S_x=0$ no ratchet
effect is possible. Accordingly, using \refm{Nnum} we find
$\overline{\mitt{N_z}}=0$ in regimes 1 and 3. In regime 2 we have
$S_x\neq 0$ which allows non-zero values of
$\overline{\mitt{N_z}}$. This is indeed what we find with, however,
two peculiarities. First, the values for $\overline{\mitt{N_z}}$ are
extremely small as long as $K/D$ is still near to the value 2
separating region 2 from region 1 (cf. Fig.~\ref{tourqehx0}). Second, 
for values of $K/D$ around the transition from region 2 to region 3 
the system apparently relaxes very slowly to its
asymptotic behaviour and it becomes very difficult to extract reliable
values for $\overline{\mitt{N_z}}$ from the numerics. The dependence
of $\overline{\mitt{N_z}}$ on the ratio $K/D$ as obtained numerically
is summarized in Fig.~\ref{tourqehx0}. 

\section{Effective Field approximation}
\begin{figure}
\begin{center}
\includegraphics[width=0.6\textwidth]{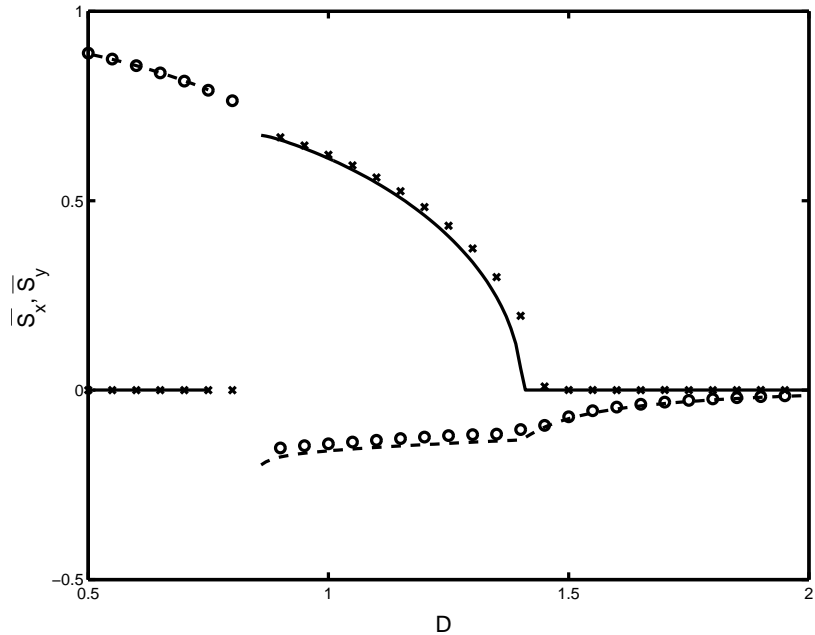}
\end{center}
\caption{Time averaged values of the $x$-(solid lines and crosses) and
  $y$-component (dashed lines and circles) of the average orientation
  $\vS$ as function of the noise strength $D$. The other parameter
  values are as in Fig.~\ref{regimes}. Lines are numerically exact
  results from the Fokker-Planck equation \refm{NFPE} whereas symbols
  denote the results obtained from the numerical solution of the
  effective field equations \refm{EFE1} and \refm{EFE2}.}
\label{exundey} 
\end{figure}

In order to obtain an improved understanding of the numerical results
obtained in the previous section we now turn to an approximate
analytical analysis of the system. A valuable tool in this
respect is the effective field approximation for ferrofluids as
extensively reviewed in \cite{RaikherS94}. To apply this approximation
in the present context the following steps are performed. From the
central equation \refm{NFPE} we derive an exact equation for the
time evolution of $\mitt{\vu}$ of the form
\bq
    \di_t \mitt{\vu} + D \mitt{\vu} = -\mitt{ \vu ( \vu  \cdot
      \vH_{mf}) } + \vH_{mf} \label{equ:efe1}\; . 
\eq
where $\vH_{mf}$ is determined by \refm{mfhfeld}. 
In order to decouple the higher moment on the r.h.s. of
\refm{equ:efe1} we approximate the unknown distribution $P(\phi,t)$ by
an {\em instantaneous equilibrium distribution} of the form \refm{Peq}
corresponding to a so far undetermined {\em effective} magnetic field
$\vH_e(t)$  
\bq \label{Peff}
    P(\phi,t) \simeq \frac{1}{\displaystyle 2 \pi
      I_0(\frac{H_e}{D})}\exp \left( 
      \frac{H_{e,x} \cos(\phi) + H_{e,y} \sin (\phi)}{D}
    \right) \; .
\eq
Using this distribution to calculate the averages in \refm{equ:efe1}
and using (cf. \refm{Sav})
\bq
         \vS=\Be{\frac{H_e}{D}} \frac{\vH_e}{H_e}   \label{morvec}
\eq
the following evolution equation for the effective field $\vH_e(t)$
may be derived
\begin{equation}
 \di_t \left[ \Be{\frac{H_e}{D}} \frac{\vH_e}{H_e} \right]  =  
   - \frac{H_e-2D \Be{\frac{He}{D}}}{H_e^3} \vH_e \times (\vH_e \times
   \vH_{mf})- \frac{ \Be{\frac{H_e}{D}}}{H_e}
   \left(\vH_e-\vH_{mf} \right)\; .
\end{equation}
With the help of 
\bq
    \vH_{mf}=\vH + K \Be{\frac{H_e}{D}}\frac{\vH_e}{H_e}\; 
    \label{selbstkonsistenz} 
\eq
which follows from \refm{mfhfeld} and \refm{morvec} and
representing $\vH_e$ by modulus and phase according to 
$\boldsymbol{\vH_e}=H_e( \cos \varphi, \sin \varphi)$ we find the following
closed set of nonlinear ordinary differential equations for the time
evolution of the effective field 
\begin{align}\label{EFE1}
  \di_t \big(\frac{H_e}{D}\big) &= - \frac{ \displaystyle
    \frac{D}{H_e} \Be{\frac{H_e}{D}}} 
    {\displaystyle 1-\frac{D}{H_e}
      \Be{\frac{H_e}{D}}-B^2\left({\frac{H_e}{D}}\right)} 
  \left( H_e  - K \Be{ \frac{H_e}{D} }-H_y(t)\sin \varphi \right) \\ 
    \di_t \phi &= \frac{\displaystyle 1- \frac{D}{H_e} \Be{\frac{H_e}{D}}}
             {\displaystyle \Be{\frac{H_e}{D}}} H_y(t) \cos \varphi \; .
    \label{EFE2} 
\end{align}
These equations cannot be solved analytically, however their numerical
solution is much simpler than the numerical solution of the
Fokker-Planck equation \refm{NFPE}. As shown in Fig.~\ref{exundey} the
results provide rather accurate estimates for the relevant quantities. 


\section{Different manifestations of the ratchet effect}
With the help of the effective field approximation it is possible 
to gain a more intuitive understanding of the existence of 
different regimes of the system behaviour as found in section
\ref{NUM}. Moreover, their stability as well as the operation of the
ratchet effect in the different regimes may be elucidated. 

Using the evolution equation for the effective field
\refm{EFE1},\refm{EFE2} and the relation \refm{morvec} between the average
collective orientation $\vS$ and the effective field we may write the time
evolution of $\vS$ in the form 
\begin{align}
   \di_t \vS = -\nabla V(S) + \boldsymbol{F}(\vH,\vS)\; .
   \label{gleichungorient} 
\end{align}
The r.h.s. of \refm{gleichungorient} was split into a central
potential term incorporating 
the effects of diffusion and interaction and an external force field
related to $\vH(t)$. Note that $0\leq S\leq 1$ always.

\begin{figure}
\begin{center}
\includegraphics[width=0.8\textwidth]{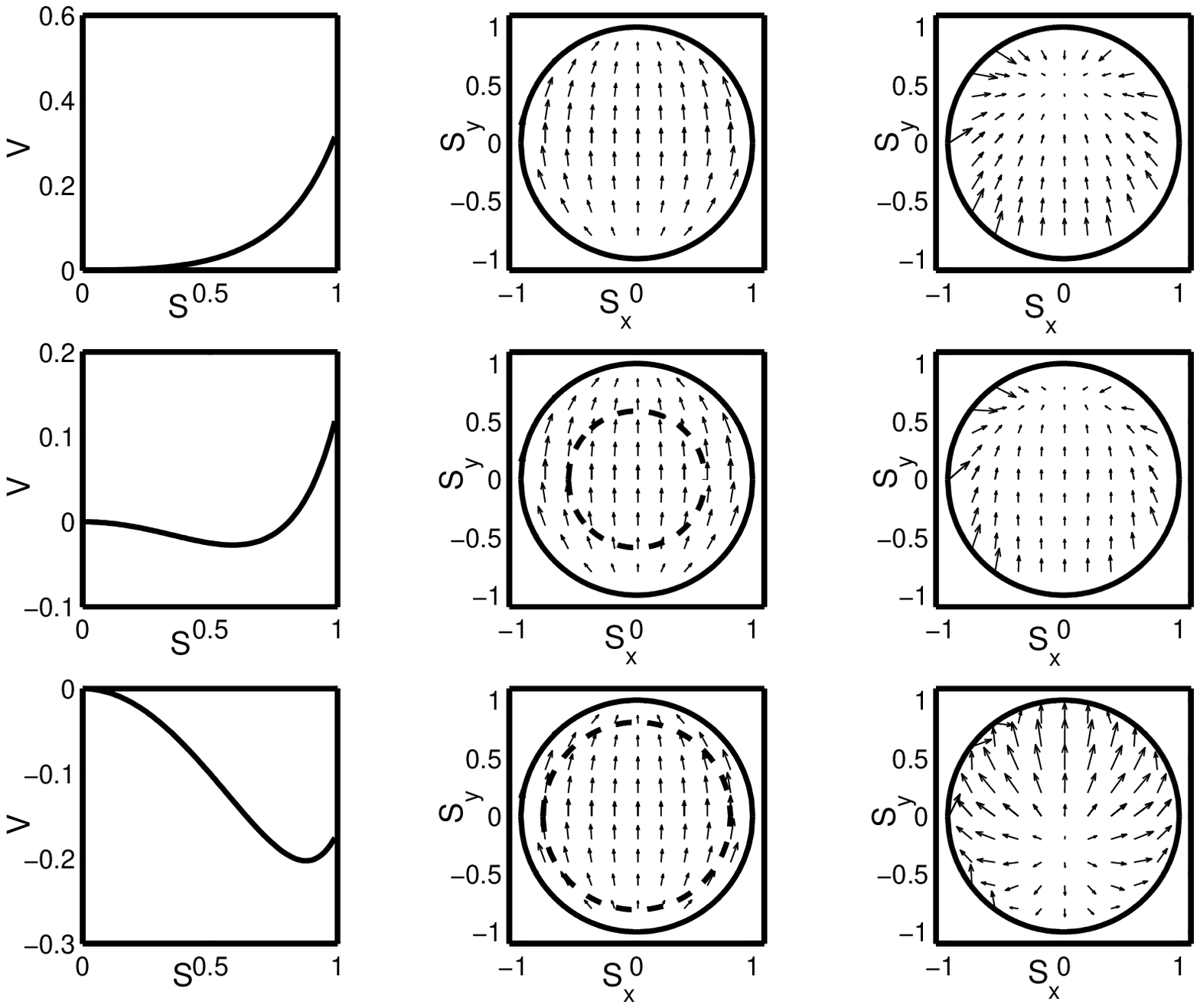}
\end{center}
\caption{Plot of the different terms contributing to the r.h.s. of 
  \refm{gleichungorient} for $H_x=0$, $K=3$ and a time $t$ for which 
  $H_y(t)=1$. The first line corresponds to $D=1.6$ (regime 1), the 
  second one to $D=1.2$ (regime 2), and the third one to $D=0.6$ 
  (regime 3). Shown are the potential $V(S)$ defined in 
  \refm{potentialorient} in the left column, the external force field 
  $\boldsymbol{F}$ defined in \refm{potentialorient} in the middle 
  column and the  total resulting force field in the right column. The
  dashed line in the middle column marks the equilibrium value $S_{eq}$.
  \label{feld}} 
\end{figure}

Let us first discuss the potential part. It is given by 
\bq
  V(S) = \frac{D}{2} S^2 - K \int_0^S dS' \frac{S'^2}{B^{-1}(S')}\;
  , \label{potentialorient}
\eq
where $B^{-1}$ denotes the inverse function of $B$ defined in
\refm{defBx}. Hence $B^{-1}$ is linear for small values of $S$ and
tends to infinity for $S\to 1$ (cf. \refm{propB1},\refm{propB2}). For
small values of $K/D$ the 
potential $V(S)$ therefore has a minimum at $S=0$ whereas for $K/D>2$
it develops a non-trivial minimum at $S_{eq}>0$ 
(cf. Fig.~\ref{feld}, left column). From \refm{selbstkonsistenz} and
\refm{EFE2} the corresponding equilibrium value of the effective field
satisfying 
\bq \label{ggmefe0}
   \Be{\frac{H_{eq}}{D}} = S_{eq}
\eq
is determined by 
\bq \label{ggmefe}
      H_{eq}=K \Be{\frac{H_{eq}}{D}}\; .
\eq
We note that with increasing $K$ also $S_{eq}$ increases (see
Fig.~\ref{feld}). In the absence of an external magnetic field, $H=0$,
the stationary solution $S_{eq}$ of the effective field equations
coincides with the {\em exact} equilibrium solution of the
Fokker-Planck equation \refm{NFPE}. 

For the external force field we get 
\bq
      \boldsymbol{F}(t)= \frac{S}{B^{-1}(S)} H_y(t) \sin \varphi
      \; \e_{r} + \left( 1- \frac{ S}{B^{-1}(S)}\right) H_y(t)
      \cos \varphi  \;\e_{\varphi} \label{kraftorient} \; .
\eq
For small values of $S$ this field is parallel to the external
magnetic field $\vH$. For $S\to 1$ on the other hand the radial
component in $\boldsymbol{F}$ is suppressed and only the tangential 
term survives (cf. Fig.~\ref{feld}, middle column). This is quite
intuitive since for $S\to 1$ all magnetic moments are aligned and
therefore the external field induces identical changes to their
orientations giving rise to a change in the {\em orientation} of $\vS$
only.  

For small values of the external magnetic field $S$ will only slightly
differ from its equilibrium value $S_{eq}$. We will hence find
$S\simeq 0$ for small values of $K/D$ which was characteristic of
regime 1. Larger values of $K/D$ result in values
of $S$ differing substantially from zero. From \refm{kraftorient} it
is then clear, that $\varphi=\pi/2$ is a possible steady state since
in this case $\boldsymbol{F}$ is parallel to $\e_y$ and there is hence
no systematic force deflecting $\vS$ from the orientation along the
$y$-axis. This situation corresponds to regime 3. Finally, there is
also a steady state solution with $s>0$ and $\varphi\simeq 0$ (regime 2),
which is, however, not obvious from \refm{kraftorient}.

We will now give a simple argument for the transition between
regime 2 and regime 3 which allows to derive an estimate for the
threshold value of $K/D$ separating these two regimes. Since the
potential part in \refm{gleichungorient} only influences the modulus
of $S$ the time evolution of the angle $\varphi$ is solely given by 
\bq \label{anderungphi}
      \di_t \varphi = \frac{1}{S} F_{\varphi}\; .
\eq
Assume now that at some given point of time we have
$0<\varphi(t)<\pi/2$. As long as $H_y(t)$ is positive (as in
Fig.~\ref{feld}) $\vS$ will be pushed in positive $y$-direction
(cf. Fig.~\ref{feld}, right column) implying a slight increase of $S$
beyond the value $S_{eq}$. Similarly, when somewhat later $H_y(t)$
changes sign and becomes negative $\vS$ will be pushed in negative
$y$-direction and $S$ will slightly decrease. Simultaneously $\varphi$
changes according to \refm{anderungphi}, i.e. increases when
$H_y(t)>0$ and decreases when $H_y(t)<0$. However, the {\em rate} of
change depends on $S(t)$. A rough estimate on whether the increase of
$\varphi$ dominates over its decrease or vice versa may be obtained from
the quantity 
\bq \label{bedingungsgleichung} 
      J(\varphi)=\left. \frac{\partial}{\partial S} \left(
          \frac{F_{\varphi}}{H_y(t)S} \right)  \right|_{S=S_{eq }} \; .
\eq
If $J(\varphi)<0$ the decrease of $\varphi$ will dominate, if
$J(\varphi)>0$ its increase. Remarkably, although $J$ depends on
$\varphi$ its {\em sign} depends only on the ratio $K/D$. Hence either 
$\varphi$ decreases all the time until it reaches $\varphi=0$ and we are in
regime 2 or it increases up to $\varphi=\pi/2$ corresponding to regime
3. An explicit calculation yields 
\bq
        J=\frac{K^2 (H_{eq}^2+DK+D^2-K^2)}{H_{eq}^2 (K^2-DK-H_{eq}^2)}
\eq
As shown in the appendix, the denominator is positive for all values
$K/D>2$. The condition for the stability of regime 2 $J<0$ is hence 
\bq\label{intbed2}
      H_{eq}^2 + KD + D^2-K^2<0\; ,
\eq
which together with \refm{ggmefe} yields that regime 2 is stable if 
\bq
    \Be{\sqrt{\frac{K^2}{D^2}-\frac{K}{D}-1}}<
   \sqrt{1-\frac{D^2}{K^2}-\frac{D}{K}} \label{intbed}  
\eq
or equivalently
\bq
      \frac{K}{D} < 3.75...\; .
\eq

As we will see in subsections \ref{sec:regime2} and \ref{sec:regime3}
below a more systematic investigation of the stability of regimes 2
and 3 will give rise to the same result (cf. \refm{stab2},\refm{h8}). 


\subsection{Regime 1}
\label{sec:regime1}

Regime 1, characterized by $K/D<2$, is the most relevant for
ferrofluids. There is no spontaneous collective orientation,
i.e. $\vS=0$, and 
accordingly we must have $H_x \neq 0$ in order to find a noise induced
rotation. The situation is hence similar to the case without
interaction, nevertheless we will show that the interactions bring
about a strongly reinforced ratchet effect. 

\begin{figure}
\begin{center}
\includegraphics[width=0.6\textwidth]{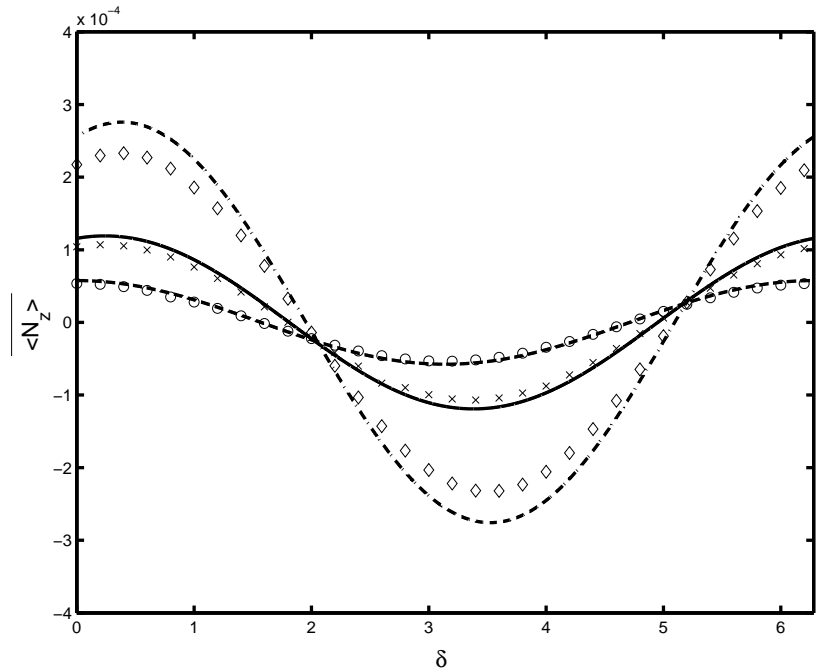}
\end{center}
\caption{The time averaged torque in regime 1 as function of the phase
  $\delta$ in the time dependence \refm{foft} for different values of the
  coupling strength $K$. The symbols are results from the numerical
  solution of the Fokker-Planck equation \refm{NFPE}, the lines show
  the results of the perturbation theory outlined above. The parameter
  values are  $H_x=\alpha=\beta=1/ \sqrt{6}$, $D=1.6$, and $K=0$
  (dashed line and circles), $K=0.5$ (solid line and crosses) and
  $K=1$ (dashed-dotted line and diamonds) respectively.}
 \label{tourqehx} 
\end{figure}

To obtain a quantitative estimate of this reinforcement we use a
variant of the perturbation theory for small values of the external
field introduced in \cite{EngelR04}. To keep track of the different
orders in the expansion it is convenient to use the
rescaling  
\bq \label{Hpert}
\vH \rightarrow \epsilon \vH \; .
\eq 
We then solve the Fokker-Planck equation \refm{NFPE} using the
expansion 
\bq
        P=P^{(0)} + \epsilon P^{(1)} + \epsilon^2 P^{(2)}+ \epsilon^3
        P^{(3)}+... 
\eq
with the unperturbed solution given by 
\bq
        P^{(0)} = \frac{1}{2 \pi}\; .
\eq
Using \refm{Fourier} this ansatz gives rise to a similar expansion for
the coefficients $a_n$ with $n>0$ 
\bq
        a_n=\epsilon a_n^{(1)} + \epsilon^2 a_n^{(2)}+ ...
\eq
whereas $a_0$ is to all orders in $\epsilon$ fixed by the
normalization condition to 
\bq
        a_0=\frac{1}{2\pi} \; .
\eq
The peculiarity of the first regime is that $S\to 0$ when $H\to 0$.
We may therefore consistently employ a similar expansion for the
collective orientation $\vS$, 
\bq
        \vS = \epsilon \vS^{1}+\epsilon^2 \vS^{2}+...\; .
\eq
The first non-zero result for $\overline{\mitt{N_z}}$ is obtained in
fourth order in $\epsilon$. From \refm{Nnum} we infer that we hence
need $a_1$ up to third order. Similar to \cite{EngelR04} we find to
first order in $\epsilon$ 
\bq
    \left( \di_t+D- \frac{K}{2} \right) a_1^{(1)} = \frac{H_x-i
      H_y(t)}{4 \pi}\; ,
\eq
and to second order
\begin{align}\label{h3}
  \left( \di_t +D - \frac{K}{2} \right) a_1^{(2)} & = 0 \\
  \left( \di_t +4D \right) a_2^{(2)}&=(H_x-i H_y) a_1^{(1)}+2 \pi K
  a_1^{(1)} a_1^{(1)}   
\end{align}
In the first regime we have $D-K/2>0$. From
\refm{h3} we then get $a_1^{(2)}(t)\to 0$ for $t \to \infty$. 
To third order we find
\bq
  \left( \di_t + D -\frac{K}{2} \right)a_1^{(3)} = -\frac{H_x + i
          H_y}{2} a_2^{(2)}- \pi K a_{-1}^{(1)} a_2^{(2)} \; .
\eq
Using the special time dependence \refm{foft} and solving for the
asymptotic behaviour of $a_1^{(1)}$ and $a_1^{(3)}$ we find for
the torque \refm{Nnum} after some algebra
\begin{align}
  \overline{\mitt{N_z}} &=6 H_x \alpha^2 \beta \left\{ \frac{\left(128
        D^5 K -64 D^4 K^2 + 40 D^3 K - 20 D^2 K^2 + 8 DK - 2
        K^2\right) \sin \delta}{
      (16D^2+1)(4D^2-4DK+K^2+4)^2(2D-K)(4D^2-4DK+K^2+16)(4D^2+1)}
  \right.\nonumber\\  
        &+\left. \frac{\left( 184 D^5 + 68 D^4 K + 58 D^3 K^2 + 200
              D^3 + 8 D^2 K - 3 D^2 K^3 + 16 D + 6 D K^2 + 4K\right)
   \cos\delta}{(16D^2+1)(4D^2-4DK+K^2+4)^2(2D-K)
    (4D^2-4DK+K^2+16)(4D^2+1)} \right\}   \; .
\end{align}
For $K=0$ this expression simplifies to 
\cite{EngelR04} 
\bq
   \overline{\mitt{N_z}}= \frac{ 3 H_x \alpha^2 \beta}{8}\,
    \frac{(23 D^2+2) \cos\delta }{(4 D^2 +1)(D^2+4)(D^2+1)(16 D^2+1) } \; ,
\eq
which coincides with the result found in \cite{EngelR04} when
specialized to the $x$-$y$ plane, i.e., to $\theta\equiv \pi/2$. 
Fig.~\ref{tourqehx} shows the torque as function of the phase angle
$\delta$ for different values of the interaction
strength $K$. It is clearly seen that although the interaction is in
regime 1 too weak to induce qualitative changes in the behaviour the
value of the noise-induced torque is greatly enhanced by the
interaction between the particles. 


\subsection{Regime 2}
\label{sec:regime2}
In regime 2 the interactions between the particles give rise to a
non-zero value of the $x$-component $S_x$ of the collective
orientation $\vS$. This breaks the symmetry $x\mapsto -x$ even in the 
absence of an $x$-component of the external field, $H_x=0$, and
results in an interaction sustained ratchet effect in the system. In
the present subsection we derive an approximate solution of the
effective field equations \refm{EFE1},\refm{EFE2} characteristic for
this regime and discuss its stability. 

To obtain an approximate analytical solution valid in the long-time
limit we resort again to perturbation theory and assume that $H$ is
much smaller than the equilibrium value $H_{eq}$ of the effective
field $H_e$. This 
assumption can easily be  met if we are sufficiently far from the
boundary of regime 2, $K/D=2$, at which $H_{eq}$ tends to zero. To
formally organize the perturbation expansion it is again convenient to
employ the rescaling \refm{Hpert} and to write the solution of the
effective field equations \refm{EFE1},\refm{EFE2} as power series in
$\epsilon$ 
\begin{align}\label{pertHe}
     H_e &= H_e^{(0)}+\epsilon H_e^{(1)} + \epsilon^2 H_e^{(2)} + ...\\
     \varphi &=\varphi^{(0)} + \epsilon \varphi^{(1)}+\epsilon^2 \varphi^{(2)} +
     ... \; ,
\end{align} 
where $H_e^{(0)}=H_{eq}$ and $\varphi^{(0)}$ is not specified at this
stage. Using these ans\"atze in \refm{EFE1},\refm{EFE2} and 
matching powers of $\epsilon$ we find to first order 
\begin{align}\label{He1}
   \di_t H_e^{(1)} &= D \frac{K^2 - H_{eq}^2 - 2 KD}{K^2-H_{eq}^2-KD} H_e^{(1)} + D^2 K \frac{\sin \varphi_0}{K^2-H_{eq}^2-KD}H_y(t) \\ 
    \di_t \varphi^{(1)}&=\frac{(K-D) \cos(\varphi_0)}{H_{eq} } H_y(t)\; .
\end{align}
These equations can be solved for arbitrary $\varphi_0$. The second order
equations are rather long and will not be displayed in full
generality. However, as typical for degenerate perturbation theory,
the second order equations may contain secular terms. In the present
case this happens unless either $\varphi_0=\pm \pi/2$ or 
$\int_0^{2\pi} dt \, H_y(t) H_e^{(1)}(t)=0$. The first condition
corresponds to regime 3 to be discussed in the next subsection. The
second one is equivalent to $\varphi_0=0$ as follows from \refm{He1} and
is therefore the condition appropriate for the present investigation
of regime 2. Using $\varphi_0=0$ and the specific form \refm{foft} of the
time dependence of $H_y(t)$ the first order solution tends for large
$t$ to 
\begin{align} \label{fohe}
   H_e^{(1)}&=0 \\
   \varphi^{(1)}&= \frac{K-D}{H_{eq}} \left( \alpha \sin(t) -
     \frac{\beta}{2} \cos(2t) \right) + \varphi_m \label{fophi}  \; ,
\end{align}
where $\varphi_m$ is an integration constant. Using these results the
second order equations greatly simplify and take the form 
\begin{align}
        \di_t H_e^{(2)}&=-A H_e^{(2)} + B H_y(t) \varphi^{(1)} \\
        \di_t \varphi^{(2)} &=0 \label{sohe}\; ,
\end{align}
with 
\begin{align}\label{defA}
        A &= D\frac{K^2-2KD-H_{eq}^2}{H_{eq}^2+KD-K^2} \\
        B &= -\frac{D^2 K}{H_{eq}^2+KD-K^2}\; .\label{defB}
\end{align}   
Hence $\varphi^{(2)}$ is a constant and $H_e^{(2)}$ may easily be
calculated for all values of $\varphi_m$. However, this solution produces
secular terms in the third order equations unless
\bq
        \int_0^{2\pi} dt H_y(t) H_e^{(2)}(t)=0
\eq
implying
\bq\label{resphim}
     \varphi_m = \frac{3 \beta \alpha^2 (A^2+2)}{4(\alpha^2 A^2 + \beta^2
       A^2 + 4 \alpha^2 + \beta^2)} \frac{K-D}{H_{eq}} \; .
\eq
This result for $\varphi_m$ completes the first order solution of the
effective field equations. Higher orders may be obtained along the
same lines but will not be considered here. 

\begin{figure}
\begin{center}
\includegraphics[width=0.6\textwidth]{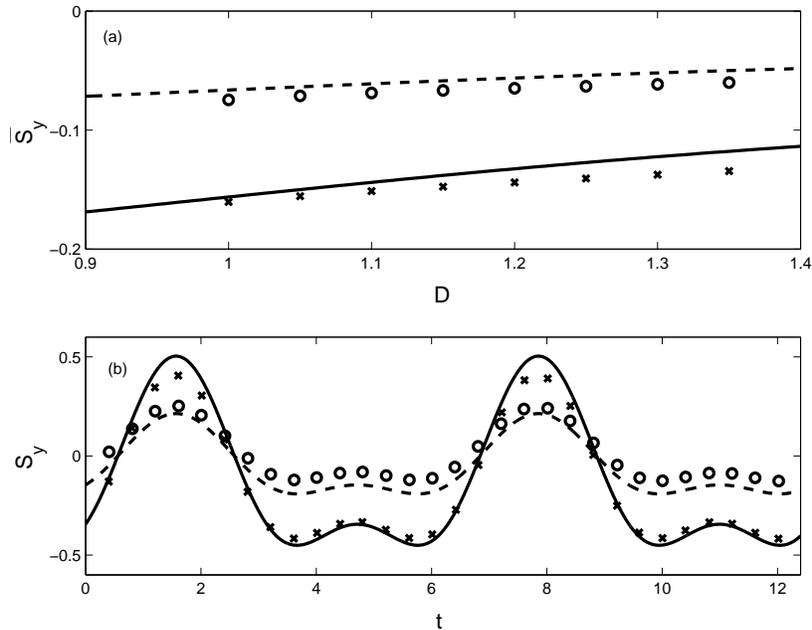}
\end{center}
\caption{(a) Time average of the $y$-component $S_y$ of the collective
  orientation $\vS$ in regime 2 as function of the noise strength $D$ for
  interaction parameter $K=3$. Symbols are numerical results from the 
  Fokker-Planck equation, lines show perturbative results from
  the effective field equations. The solid line and the crosses are
  for $\alpha=\beta=1/\sqrt{2}$, the dashed line and the circles are
  for $\alpha=\beta=0.3$. (b) Time evolution of the $y$-component of
  the orientation vector $\vS$ in regime 2. The noise intensity is
  $D=1.2$, all other parameter values are as in (a).     
  \label{numvsanr2}}
\end{figure}

For the collective orientation we then obtain to first order in the
external field 
\begin{align}
        S_x &= \frac{H_{eq}}{K} \\
        S_y &= \frac{H_{eq}}{K} \varphi^{(1)}\; .
\end{align}
From \refm{fophi} we hence infer that the time average of $S_y$ is directly
related to $\varphi_m$, 
\bq
        \overline{S_y}=\frac{H_{eq}}{K} \varphi_m \; .
\eq
Fig.~\ref{numvsanr2} compares the first order results obtained above
with numerically exact results from the Fokker-Planck equation
\refm{NFPE} for different parameter values. The agreement is rather
good showing that both parameter sets for $\alpha$ and $\beta$
correspond to external field strengths well within the region of
validity of the perturbation theory.  

We now turn to the investigation of the stability of the first order
solution given by \refm{fohe},\refm{fophi}, and \refm{resphim}. To this end
we add small deviations $\delta H_e$ and $\delta \varphi$ to this
solution and study their time evolution by linearizing \refm{EFE1},
\refm{EFE2} around $H_e^{(1)}(t), \varphi^{(1)}(t)$ \cite{Jordan99}. 
In this way we find 
\bq \label{pertsystem}
        \di_t \left( \begin{matrix} \delta H_e \cr \delta \varphi
          \end{matrix} \right) = \left( \begin{matrix} a_{11} &
            -a_{12} H_y(t) \cr a_{21} H_y(t) & 0 \end{matrix} \right)
        \left( \begin{matrix} \delta H_e \cr \delta \varphi
          \end{matrix}\right)     \; ,
\eq 
where 
\begin{align}
    a_{11}&=\frac{H_{eq}^2+2DK-K^2}{H_{eq}^2+DK-K^2}=-A \label{defa11}\\ 
    a_{12}&=\frac{H_{eq}D^2K}{H_{eq}^2+DK-K^2}=-B \label{defa12}\\
    a_{21}&=\frac{KD+H_{eq}^2+D^2-K^2}{H_{eq}^2 D}\label{defa21}\; .
\end{align}
We are interested in the stability of the fixed point $\delta
H_e=\delta \varphi=0 $ of this system. Because of the periodic time
dependence of the coefficient matrix in \refm{pertsystem} a general
analysis requires Floquet theory \cite{Jordan99}. However, the special
form of \refm{pertsystem} allows us to use a more direct method. 

We first note that one can show (see appendix) $a_{11}<0$
within regime 2. We next define the function  
\bq
      V(\delta H_e , \delta \varphi) =  \frac{a_{21}}{2} 
        ( a_{21} \delta H_e^2 + a_{12} \delta \varphi^2)\; . 
\eq
For its time derivative we find 
\bq
        \frac{dV}{dt} = a_{21}^2 a_{11} H_e^2  < 0 \label{lyp1}\; .
\eq
Moreover if 
\bq\label{stab2}
        a_{21} a_{12} >0 
\eq
we have for all values of $\delta H_e$ and $\delta \varphi$ 
\bq
        V(\delta H_e, \delta \varphi) \geq 0  \label{lyp2}
\eq
with equality being valid only for $\delta H_e=\delta \varphi=0$. 
$V(\delta H_e , \delta \varphi)$ is hence a Lyapunov function
\cite{Glendinning94} of the system \refm{pertsystem} and the point 
$\delta H_e=\delta \varphi=0$ is stable as long as \refm{stab2} holds. 
From \refm{defa12} one finds $a_{12}<0$ (see appendix) within
regime 2 and hence the 
first order solution $H_e^{(1)}(t), \varphi^{(1)}(t)$ is stable if
$a_{21}<0$. From \refm{defa21} it is seen that this condition is
identical with \refm{intbed2}. We hence find again that regime 2 is
stable if 
\bq
               2< \frac{K}{D} \lesssim 3.75 \; .
\eq 
Note that the entire stability analysis performed above is valid
only to first order in the external field strength.

\subsection{Regime 3}
\label{sec:regime3}
\begin{figure}
\begin{center}
\includegraphics[width=0.6\textwidth]{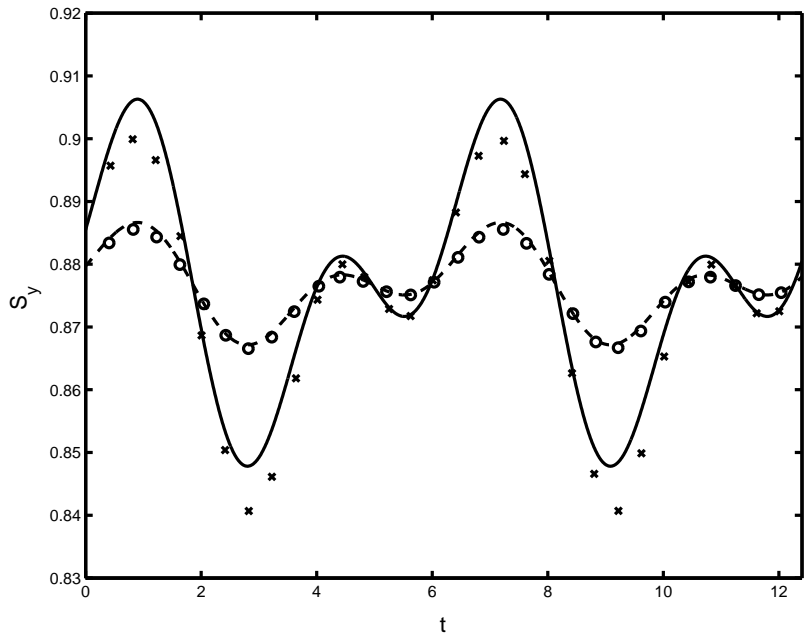}
\end{center}
\caption{Time evolution of the $y$ component of the orientation
  vector $\vS$ in regime 3. Symbols are numerical results from the 
  Fokker-Planck equation, lines show perturbative results from
  the effective field equations. The parameters are $D=0.6$ and $K=3$,
  as well as $\alpha=\beta=0.3$ (crosses and full line) and $\alpha=\beta=0.1$
  (circles and dashed line)respectively.
  \label{reg3aprox}}
\end{figure}

Similarly to the previous subsection we are now using the effective 
field equations \refm{EFE1},\refm{EFE2} to derive an approximate
analytical solution typical for regime 3 and to discuss its stability.
In regime 3 we have $\varphi=\pi/2$ (the case $\varphi=-\pi/2$ can be
dealt with similarly) and we find from \refm{EFE1},\refm{EFE2}
\begin{align}
  \di_t \big(\frac{H_e}{D}\big) &= - 
    \frac{ \displaystyle \frac{D}{H_e} \Be{\frac{H_e}{D}}}
    {\displaystyle 1-\frac{D}{H_e}
      \Be{\frac{H_e}{D}}-B^2\left({\frac{H_e}{D}}\right)} 
  \left( H_e  - K \Be{ \frac{H_e}{D} }-H_y(t)\right) \\ 
    \di_t \varphi &= 0 \; .
\end{align}
From the second equation we get immediately $\varphi\equiv \pi/2$. The
first equation determines $H_e(t)$ and cannot be solved in closed
form. Focusing again on the situation of small external fields we use
\refm{pertHe} and obtain for the first order term the equation
\bq
        \di_t H_e^{(1)}= -A H_e^{(1)} + B H_y(t)
\eq
with $A$ and $B$ defined by \refm{defA} and \refm{defB} respectively. 
The solution is given by 
\bq\label{h5}
        H_e^{(1)} =B\, e^{-At} \int dt \, H_y(t) \;e^{A t}  \; .
\eq
For the special time dependence \refm{foft} we find for large $t$
\bq
        H_e^{(1)}=B \alpha \frac{A \cos(t)+\sin(t) }{A^2+1} + B \beta
        \frac{A \sin(2t)-\cos(2t)}{A^2+4} \; .
\eq
In Fig.~\ref{reg3aprox} this result is compared with the solution of the
Fokker-Planck equation showing again good agreement for the parameter
values chosen. 

We now turn to the investigation of the stability of the first order
solution. In view of the form of the potential $V(S)$
in regime 3 (see Fig.~\ref{feld} first column, last line) it is
reasonable to assume that the state will be destabilized by
perturbations in $\varphi$. Restricting ourselves to this case we use
$H_e(t)$ as determined above, set $\varphi(t) = \pi/2 + \delta
\varphi(t)$, and determine the linearized time  
evolution of $\delta \varphi(t)$. From \refm{EFE2} we get
\bq
\di_t \delta\varphi = -\frac{\displaystyle 1- \frac{D}{H_e}
  \Be{\frac{H_e}{D}}} 
  {\displaystyle\Be{\frac{H_e}{D}}} H_y(t) \, \delta\varphi 
\eq
with solution 
\bq
    \delta \varphi(t) = \delta \varphi_0 \exp\left(-\int_0^t dt' 
           \frac{1- \frac{D}{H_e} \Be{\frac{H_e}{D}}}
             {\Be{\frac{H_e}{D}}}H_y \right)\label{stoerung}\; .
\eq
Since both $H_y$ and $H_e$ are $2 \pi$-periodic functions of time this
solution may be written in the form 
\bq
        \delta \varphi(t)=\delta \varphi_0 \; e^{-a_0 t} \; f(t)\; ,
\eq
where $f(t)$ is also $2\pi$-periodic and $a_0$ is the constant term in
the Fourier expansion of the exponent in \refm{stoerung} and therefore
given by  
\bq
    a_0 = \frac{1}{2 \pi} \int_0^{2\pi} dt \, 
  \frac{1- \frac{D}{H_e} \Be{\frac{H_e}{D}}}
     {\Be{\frac{H_e}{D}}}H_y \; .
\eq
Using the first order solution $H_e(t)=H_{eq}+H_e^{(1)}(t)$ we find 
\bq
        a_0= \frac{(D K+D^2-K^2+H_{eq}^2)}{2 \pi H_{eq}^2 D}
    \int_0^{2 \pi} dt' \, H_e^{(1)}(t')\, H_y(t') \; .\label{regimestab} 
\eq
The integral in this expression is in regime 2 and in regime 3 always positive
as can be shown by expanding $H_y(t)$ in a Fourier series
\bq
        H_y = \sum_{n=-\infty}^{\infty} b_n e^{int}
\eq
and using \refm{h5}. We then find 
\bq
        \frac{1}{2\pi} \int_0^{2 \pi} dt H_e^{(1)}(t)\, H_y(t) =
        \sum_{n=0}^{\infty} \frac{B}{n^2+A^2}\, |b_n|^2 \geq 0 \; ,    
\eq
since $B>0$ as demonstrated in the appendix. 
Hence the sign of $a_0$ as given by \refm{regimestab} depends solely
on the prefactor and the solution is unstable if 
\bq\label{h8}
       D K+D^2-K^2+H_{eq}^2<0\; .   
\eq
This condition is again identical with \refm{intbed} and we hence find
that regime 3 becomes unstable at the same value of $K/D$ at which
regime 2 becomes stable.

\section{Conclusions}

In the present paper we have theoretically analyzed the influence of
particle-particle interactions of the mean-field type on the ratchet
effect in ferrofluids. We have used the simplified mean-field model
because the details of the realistic dipole-dipole and hydrodynamic
interactions are too complicated for a general discussion. On the
other hand, several qualitative effects which may be expected to show
up in case of these realistic interactions can already be discussed
within the mean-field approach. 

Quite generally we have found that interactions that favour the parallel
alignment of the magnetizations of the ferrofluid particles reinforce
the ratchet effect. In the present situation this means that the
magnetic torque per volume of the ferrofluid is enhanced. Although
intuitive this result in by no means trivial since the coupling
between the ferromagnetic grains is also likely to reduce their
orientational fluctuations which are the driving force of the ratchet
effect. 

Moreover we have shown that for sufficiently strong coupling
giving rise to a spontaneous symmetry breaking with respect to the
collective orientation of the grains the ratchet effect may be brought
about for situations in which it could not operate in the absence of
interactions. In the present system this happens for a purely
oscillating external magnetic field without constant component in the
$x$-direction. In this case and without interactions no ratchet effect
is possible due to symmetry reasons whereas a spontaneous breaking of
the relevant symmetry caused by the interactions may induce a
rectification of fluctuations. 

For even stronger coupling strength between the particles this 
self-sustained ratchet effect again disappears. The particle 
orientations are now highly aligned and closely follow the direction
of the external field. Fluctuations are therefore suppressed and no
rectification is possible anymore. 

Our analysis builds on the non-linear Fokker-Planck equation
\refm{NFPE} for the collective orientation \refm{sundu} of the
ferrofluid particles. A first insight into the behaviour of the system
is gained from a numerical solution of this equation. We then use a
variant of the effective field method, which is a well-known tool in
the theory of ferrofluids, to get approximate analytical results for
the transitions between and stability of the different regimes of
operation of the ferrofluid ratchet. 

The central parameter
distinguishing the different regimes is the ratio between the
dimensionless coupling constant $K$ and the intensity of 
the fluctuations $D$. For $K/D<2$ no self-sustained ratchet effect
is possible since the interactions are too weak to induce a
spontaneous symmetry breaking. For $2<K/D<3.75$ a self-sustained
ratchet effect may be observed with a wide spectrum of values for the
transferred angular momentum. For $3.75<K/D$ the self-sustained
ratchet effect disappears again due to an instability in the steady
state solution for the collective orientation of the particles. 

\acknowledgments
We would like to thank Dirk Rannacher for clarifying discussions.

\appendix
\section{}
In this appendix we determine the signs of some expressions needed
to determine the stability of regimes 2 and 3. In these regimes we
have $K/D>2$ the value of $S_{eq}$ is determined by the non-trivial
minimum of the potential $V(S)$. From \refm{potentialorient} we
find by differentiation 
\bq
  \frac{D}{K}=\frac{S_{eq}}{B^{-1}(S_{eq})}
\eq
verifying that $S_{eq}$ depends only on the ratio $K/D$. Using the
properties of the function $B(x)$ defined in \refm{defBx} one can show
that for all $S\in [0,1]$  
\bq
  1-\frac{S}{B^{-1}(S)}\geq S^2\; .
\eq
This in turn implies 
\bq
        K^2-D K- K^2 S_{eq}^2 \geq 0
\eq
and using \refm{ggmefe0} we find 
\bq
        K^2-DK-H_{eq}^2\geq 0\; .
\eq
Due to \refm{defB} and \refm{defa12} this immediately implies
\bq
        B=-a_{12}=-\frac{D^2 K}{H_{eq}^2+KD-K^2} \geq 0 \; .
\eq
Using again the properties of the function $B(x)$ one can also show 
\bq
  1-2\frac{S}{B^{-1}(S)}\leq S^2 \; .
\eq
Similar manipulations as used above then yield for $K/D$ >2  
\bq
        K^2-2 D K - H_{eq}^2 \leq 0 \; 
\eq
from which by using \refm{defA} and \refm{defa11} we find 
\bq
        A=-a_{11}=\frac{D(K^2-2KD-H_{eq}^2)}{H_{eq}^2+KD-K^2} \geq 0
        \; .
\eq

\bibliographystyle{apsrev}
\bibliography{literatur}
\end{document}